\documentstyle[11pt]{article}
\font \msb=msbm10 scaled \magstep1

\newcommand{\bR}{\mbox{\msb R} }
\newcommand{\bC}{\mbox{\msb C} }
\def\on#1#2{\mathop{\vbox{\ialign{##\crcr\noalign{\kern2pt}
$\scriptstyle{#2}$\crcr\noalign{\kern2pt\nointerlineskip}
\kern-2pt$\hfil\displaystyle{#1}\hfil$\crcr}}}\limits}
\def\qdet{{\rm det_q\,}}
\def\cc#1{{\cite{#1}}}

\def\l{{\lambda}}
\def\a{{\alpha}}
\def\b{{\beta}}
\def\g{{\gamma}}
\def\d{{\delta}}
\def\s{{\sigma}}
\def\e{{\varepsilon}}
\def\te#1{{\tilde{#1}}}
\def\SA{{\cal A}}
\def\SB{{\cal B}}
\def\SC{{\cal C}}
\def\SD{{\cal D}}
\def\SL{{\cal L}}
\def\SG{{\cal G}}
\def\tr{{\rm tr}\,}
\def\nn{ \nonumber }
\def\bq{ \begin{equation} }
\def\eq{ \end{equation} }
\def\ben{ \begin{eqnarray} }
\def\en{ \end{eqnarray} }
\def\ll{ \label }
\def\frac#1#2{{#1\over #2}}
\def\dfrac#1#2{{\displaystyle{#1\over#2}}}
\newtheorem{theorem}{Theorem}
\newtheorem{lemma}{Lemma}

\begin{document}
\title{
\begin{flushleft}
\sf \small Preprint ISRN-LiTH-MAT-R-95-27, 1995.
\end{flushleft}
\vskip 1cm
The Kowalewski top: a new Lax representation.}
\author{
 A.V. Tsiganov\\
{\small\it
 Department of Mathematical and Computational Physics,
 Institute of Physics,}\\
{\small\it
St.Petersburg University,
198 904,  St.Petersburg,  Russia}
}
\date{}
\maketitle

\begin{abstract}
The $2\times 2$ monodromy matrices for the Kowalewski top on
the Lie algebras $e(3)$, $so(4)$ and $so(3, 1)$ are presented.
The corresponding quadratic $R$-matrix structure is the
dynamical deformation of the standard $R$-matrix algebras.
Some tops and Toda lattices related to the Kowalewski top are
discussed.
\end{abstract}

\section{Introduction}
\ll{sec:l1}
\setcounter{equation}{0}

The main object for investigation is the Kowalewski top (KT) in
the classical and quantum mechanics. In classical mechanics the
system under consideration is a special case of motion of a
heavy rigid body with fixed point, discovered by Kowalewski in
1889 \cc{kow89}. It represents a symmetric top in a constant
homogeneous field. The principal momenta of inertia relate as
$I_1\,:I_2\,:I_3=1:1:1/2$ and the center of mass located in the
equatorial plane. In the body frame, the components of the
angular momentum $l_i$ and the Poisson vector $g_i, ~i=1, 2, 3$
be generators of the Lie algebra $e(3)$ with the Poisson
brackets:
\ben
&&\bigl\{ l_i\,, l_j\,\bigr\}=
\e_{ijk}\,l_k\,,
\qquad
\bigl\{ l_i\,, g_j\,\bigr\}=
\e_{ijk}\,g_k\,,
\nn\\ \ll{e3}\\
&&\bigl\{ g_i\,, g_j\,\bigr\}= 0\,, \qquad
i, j, k=1, 2, 3.
\nn
\en
and with the fixed Casimir operators
\[J_2=(g, g)=a;\qquad J_3=(l, g)=b.\]
 Integrals of motion for the KT are given by
\ben
J_1&=&H=\dfrac12\left(l_1^2+l_2^2+2l_3^2\right)-g_1. \nn\\
\ll{hk}\\
J_4&=&k_+k_-=(l_+^2-2 g_+)(l_-^2 - 2 g_-)\,, \nn
\en
In quantum mechanics the KT has been introduced by
Laporte \cite{lap33} (quasiclassical approach see in
\cc{kk87}).

The KT are generalaized to the Kowalewsky gyrostat
with the following hamiltonian
\bq
H=\dfrac12 (l_1^2+l_2^2+2l_3^2+2\g l_3)-g_1\,.
\ll{gyr}
\eq
Gyrostat momentum proportional $\g$ is perpendicular to the
equatorial plane. This system and their counterparts on the Lie
algebras $so(4)$ and $so(3,1)$ have been considered in
\cc{kom87}.

The main our aim is in constructing the Lax representations
and their quantum counterparts for all these systems.

As a tool for investigations we will apply linear and quadratic
$R$-matrix algebras in the quantum and classical inverse
scattering method \cite{ft87}. Let us consider an algebra
generated by noncommutative entries of the matrix $T(u)$
satisfying the famous bilinear relation (ternary relation)
\bq
R(u-v)\,{\on{T}1}(u)\,{\on{T}2}(v)=
{\on{T}2}(v)\,{\on{T}1}(u)\,R(u-v)\,, \ll{fcr}
\eq
or the quaternary relation
\bq
R(u-v)\,{\on{T}1}(u)\,S(u+v)\,{\on{T}2}(v)={\on{T}2}(v)\,S(u+v)\,
{\on{T}1}(u)\,R(u-v)\,.
\ll{re}
\eq
where we use the standard notations $\on{T}{1}(u)= T(u)\otimes
I\,, ~\on{T}{2}(v)=I\otimes T(v)$ and matrices $R(u)$ and $S(u)$
are solutions of the Yang-Baxter equation. For historical
reasons these algebras are called the algebras of monodromy
matrices. Equations (\ref{fcr}) and (\ref{re}) are called the
fundamental commutator relation (FCR) and the reflection
equations (RE) \cite{skl88}, respectively. If we consider a simple
finite-dimensional Lie algebra $\sf a$ and a $\sf a$-invariant
$R$-matrix then the algebra of monodromy matrices (\ref{fcr})
after a proper specialization gives the yangian $Y({\sf a})$
introduced by Drinfeld, while algebra of monodromy matrices
(\ref{re}) corresponds to the twisted yangians \cite{o92}.

We will consider the finite-dimensional irreducible
representations of algebras (\ref{fcr}-\ref{re}), which are
polynomials on spectral parameter $u$, only.  The entries of
monodromy matrix $T(u)$ are constructed from the generators of
yangian $t_{ij}(u)$ by the rule
\bq
T(u)=\sum_{i, j}^N t_{ij}(u)\otimes E_{ij}\in Y({\sf a})\otimes
\hbox{End}(\hbox{\bC}^N)\,, \qquad t_{ij}(u)=\sum_\alpha
t^\alpha_{ij} u^\alpha\,,
\ll{mon}
\eq
where $E_{ij}$ are the standard matrix units.
The matrix trace $t(u)$ of the matrix $T(u)$
\bq
t(u)=\tr T(u)=\sum_{k=1} T_{kk}\,(u)
\ll{tu}
\eq
yields a commutative family of operators $J_k$
\bq
[t(u), t(v)]=0\,, \quad t(u)=\sum_k \,J_k u^k, \quad u, v\in{\bC}
\eq
which are integrals of motion of some quantum integrable system.

In the classical limit algebras of monodromy matrices (\ref{fcr})
and (\ref{re}) transform into the quadratic Sklyanin algebras
\ben
\{\on{T}{1}(u), \on{T}{2}(v)\}&=&
[r(u-v), \on{T}{1}(u)\on{T}{2}(v)\,]\,,
\ll{rrpoi}\\ \nn\\
\left\{\on{T}{1}(u), \on{T}{2}(v)\right\}&=&
\left[r(u-v), \on{T}{1}(u)\on{T}{2}(v)\right]+\ll{repoi}\\
&=&\on{T}{1}(u)s(u+v)\on{T}{2}(v)-
\on{T}{2}(v)s(u+v)\on{T}{1}(u). \nn
\en
Here matrices $r(u)$ and $s(u)$ are the classical $r$-matrices,
$R(u)=1+\eta r(u)+O(\eta^2)$ by $\eta\to 0$ and similar for
matrix $S(u)$.

If one substitute $T:=1+\e L+O(\e^2)$, $r:=\e r$ and let $\e\to
0$, then we get the linear $R$-matrix algebra
\bq
\{\on{L}{1}(\l), \on{L}{2}(\mu)\}=
[r_{12}(\l, \mu), \on{L}{1}(\l)]+
[r_{21}(\l, \mu), \on{L}{2}(\mu)\,]\,,
\ll{rpoi}
\eq
(see review   \cc{rs87}).
We will start with the $4\times 4$ Lax representation for the KT
given by Reyman and Semenov-Tian-Shansky \cite{rs87}, which
obeys (\ref{rpoi}).

Following to a general scheme \cite{ft87,rs87} the Lax pairs
\bq
\dfrac{dL(\l)}{dt}+\left[L(\l), M(\l)\right]=0,
\ll{laxpair}
\eq
for the matrices $L(\l)$ and $T(u)$ are constructed by the
linear and quadratic algebras (\ref{rpoi}-\ref{repoi}). Below
we fix notations $L(\l)$ and $T(u)$ for monodromy matrices,
which satisfy to linear (\ref{rpoi}) and to quadratic
(\ref{rrpoi}) algebras of monodromy matrices, respectively.

It is well known, that in classical mechanics some Lax matrices
have been proposed for the KT. Complete their list and
discussion see in   \cc{rs87}. All these
matrices with the spectral parameter are
$N\times N$ matrices by $N> 2$. If we want
to use the method of separation of variables or to consider the
KT in quantum mechanics we can, of course, try to adjust these
matrices, for example, applying the experience by Sklyanin
\cite{skl95}.

However, the separated equations for KT in
the quasiclassical approach look like equations inherent in inverse
scattering method with quadratic $R$-matrix relations
\cite{kk87}. Since for the quantization of the KT we prefer to
construct a new monodromy matrix for the KT in the $2\times 2$
auxiliary space.  Building such matrix we will use a known matrix
in larger auxiliary space, which satisfies a linear $R$-matrix
algebra (\ref{rpoi}).

So, we want to obtain $2\times 2$ matrix with entries defined on
universal enveloping algebra from the $N\times N$ matrix with entries
belong to the loop algebra.
For this purpose we will use often the geometrical \cite{bvm87}
and algebraic \cite{kuzts89} connections
of the tops on $e(3)$ and the Toda lattices.
As a settlement,
we oblige to introduce the additive deformations of the
basic algebraic relations (\ref{fcr}-\ref{rrpoi}) and we
obtain a Lax triad for the KT
\bq
\dfrac{dL(\l)}{dt}+\left[L(\l), M(\l)\right]=N(\l),
\ll{laxtriad}
\eq
where matrix $N(\l)$ is a traceless matrix. Trace of the matrix
$L(\l)$ is a generating function of integrals of motion. This
situation is analogous to introduction of the dynamical
$r$-matrices on loop algebras \cite{ts94,krw95}, where for
description of the concrete integrable systems in given method we
were forced to expand framework of the $R$-matrix formalism to
the quality new type of $R$-matrix.


\section{Axially symmetric Neumann's system}
\ll{sec:l2}
\setcounter{equation}{0}

Let us recall some results about the Toda lattices in the
classical mechanics (see   \cc{avm80,ft87}). For the periodic
Toda lattices associated with the root system of $\SA_N$
type the corresponding Lax matrix is given by
\bq
L(\l)=\left(\begin{array}{cccccc}
 p_N  &e_{NN-1} &0   &\ldots &0 &e_{1N}\l^{-1} \\
 e_{NN-1} &p_{N-1} &e_{N-1N-2}&\ldots &0  &0 \\
 \vdots  &\ddots &\ddots &\ddots &p_2 &e_{21}\\
 e_{1N}\l^{-1} &0  &\ldots &0  &e_{21} &p_1
\end{array}\right)\,.
\ll{laxtoda}
\eq
where $e_{jn}\,= e^{(q_j-q_n)/2}$ and $(p_j\,, q_j)$ are pairs
of canonically conjugate variables.  Determinant curve of the
Lax matrix $L(\l)$ defines by the matrix $L(\l, u)=uI-L(\l)\,$.
It is a three diagonal matrix and we can  introduce the
monodromy matrix in two dimensional auxiliary space
\cite{ft87}. It reads
\ben
&&T(u)=T_N(u)T_{N-1}(u)\ldots T_1(u)=
\left(\begin{array}{cc} A& B\\
C&D\end{array}\right)(u)\,, \nn\\
&&T_k(u)=
\left(\begin{array}{cc} u-p_k& -e^{q_k}\\
e^{-q_k}&0\end{array}\right)\,.
\ll{todafact}
\en
Entries of $T(u)$ are the following functions of the
minors of $L(\l)$
\ben
A(u)&=&\left[\,\det L(\l, u)\,\right]_{\l=\infty}\,, \nn\\
B(u)&=&-e^{q_1}\,\det L^{(N, N)}(\l, u) \,, \nn\\
C(u)&=&e^{q_N}\,\det L^{(1, 1)}(\l, u) \,, \ll{etoda}\\
D(u)&=&\left[ \,\l^2\det L(\l, u)\,\right]_{\l=0}\,, \nn
\en
where $L^{(j, k)}(\l, u)$ means the matrix obtained by removing
the $j$ column and $k$ row of the matrix $L(\l, u)$. The quantum
operator $T(u)$ is constructed from the classical matrix
(\ref{etoda}). It obeys the FCR (\ref{fcr}) with the rational
$R$-matrix of $XXX$ type.  For the Toda lattices associated with
the Lie algebras of the series $\SB_n, \,\SC_n$ and $\SD_n$ a
similar correspondence has been introduced in   \cc{kuzts89a}.
Such relations can be helpful for the non-three diagonal
matrices. For instance, the monodromy matrices for the Toda
lattices associated with the root systems of $\SD_N$ type have
been constructed into thus manner.

Let us start with the following Lax matrix for the KT \cite{rs87}
\bq L(\l)=\left(\begin{array}{cccc} \g&
 \dfrac{g_-}\l&l_-&\dfrac{-g_3}\l\\
 -\dfrac{g_+}\l&-\g&\dfrac{g_3}\l&-l_+\\
 l_+&\dfrac{-g_3}\l&-2l_3+\g&-2\l-\dfrac{g_+}\l\\
 \dfrac{g_3}\l&-l_-&2\l+\dfrac{g_-}\l&2l_3-\g
 \end{array}\right)\,,
 \ll{4lax}
 \eq
using natural notations $l_\pm\,= l_1 \pm il_2$, $g_\pm = g_1 \pm
ig_2$.

The Lax representation (\ref{4lax}) has been applied by solution
of equation of motion in   \cc{rs87}.  The $3\times 3$ matrix
$L^{(1, 1)}(\l)$ constructed by $L(\l)$ (\ref{4lax}), in our
notation, describes the Goryachev-Chaplygin top \cite{bk88},
which has also a monodromy matrix on $2\times 2$ auxiliary space
satisfying the Sklyanin brackets (\ref{rrpoi}) \cite{skl85}.

Motivated by representation (\ref{etoda}) we introduce the
monodromy matrix $T_0(u)$
\ben
&&A(u)=\left[\,\det L^{(1, 1), (3, 3)}(\l, u)\,\right]_{\l=0}\,,\qquad
B(u)=\left[\,i\l \det L^{(1, 1), (3, 4)}(\l, u)\,\right]_{\l=0}\,, \nn\\
&&C(u)=\left[\,-i\l \det L^{(1, 1), (4, 3)}(\l, u)\,\right]_{\l=0}\,, \qquad
D(u)=\left[\,-\l^2 \det L^{(1, 1), (4, 4)}(\l, u)\right]_{\l=0}\,, \nn\\
\nn\\
&&T_0(u)=
\left(\begin{array}{cc} A& B\\
C&D\end{array}\right)(u)=
\left(\begin{array}{cc} u^2-2 u l_3-l_1^2-l_2^2&i (u g_+-g_3 l_+)\\
i (u g_- -g_3 l_-)&g_3^2\end{array}\right)\,.
\ll{vop}
\en
Introduction of the gyrostat parameter $\g$ is equivalent to
the shift of the spectral parameter $u\to (u-\g)$ and we put
$\g=0$ for a while.

This matrix corresponds to the axially symmetric Neumann's
system
\bq J_1=H=l_1^2+l_2^2-g_3^2\,, \qquad J_4=m=l_3\,,
 \ll{intneu} \eq
(or to the particular case of the general Lagrange top
\cite{ar77}).

The monodromy matrix $T_0(u)$ (\ref{vop}) has been introduced
before in   \cc{kuzts89} by using of an isomorphism of
universal enveloping algebras, which exist
in the one-parameter
subset of orbits ${\cal O}$ ($J_2=(g, g)=a^2$ and $J_3=(l, g)=0$) only.
At the level $J_3=(l, g)=0$ matrix $T_0(u)$ (\ref{vop}) obeys the
Sklyanin brackets (\ref{rrpoi}) with the rational $R$-matrix of
the $XXX$ type \cite{ft87}
\bq
r=\dfrac{2i}{u-v}\,
 \left(\begin{array}{cccc}
 1&0&0&0\\0&0&1&0\\0&1&0&0\\0&0&0&1\end{array}\right)=
 \dfrac{2i}{u-v}\sum_{i=1}^3 \s_i\otimes \s_i=
 \dfrac{\eta}{u-v} P\,,
 \ll{rxxx}
 \eq
where $P$ is a permutation operator of auxiliary spaces
and $\eta=2i$.

At $J_3=(l, g)=0$ the spectral invariants of the Lax matrix
$T_0(u)$ are the generating functions of the integrals of
motion (\ref{intneu}) and of the Casimir operators (\ref{e3}),
respectively:
\ben
 t_0(u)&\equiv&\tr\,T_0(u)=u^2-2uJ_4-J_1\,, \ll{t0}\\
 \Delta_0(u)&\equiv&{\det}\,T_0(u)=u^2J_2\,. \ll{d0}
 \en
So, matrix $T_0(u)$ (\ref{vop}) describes a completely
integrable system in the one-parameter subset of orbits ${\cal
O}$ ($J_2=(g, g)=a^2$ and $J_3=(l, g)=0$) in $e(3)^*$.

However, in contrast with   \cc{kuzts89},the
matrix $T_0(u)$ (\ref{vop}) was obtained by the matrix
$L(\l)$ (\ref{4lax}) defined on a whole phase space and the
Lagrange top is a complete integrable system on the
general orbits ${\cal O}$ ($J_2=(g, g)=a^2$ and $J_3=(l, g)=b$)
in $e(3)^*$ \cite{ar77}. Therefore, at the next section, we go to
investigate the matrix $T_0(u)$.

\section{Deformation of the Sklyanin brackets}
\ll{sec:l3}
\setcounter{equation}{0}

According to   \cc{ts94,krw95} we can introduce an additive
deformations of the algebras of monodromy matrices .  One
simple deformation of the matrix $T_0(u)$ has been considered
in   \cc{kuzts89,ts94}
\bq
T_1(u)=T_0(u)+
 \left(\begin{array}{cc}
 \dfrac{\mu}{g_3^2} &0\\0&0\end{array}\right)
 =T_0(u)+\mu
 \left(\begin{array}{cc}D^{-1}&0\\0&0\end{array}\right)
 \qquad \mu\in {\bR}\,.
 \ll{fneu}
 \eq
The corresponding new hamiltonian reads
$H^{new}=H^{old}+{\mu}/{g_3^2}$, where $H^{old}$ is a hamiltonian
(\ref{intneu}). Matrix $T_0(u)$ (\ref{vop}) and modified matrix
$T_1(u)$ (\ref{fneu}) obey the same quadratic $R$-matrix algebra
(\ref{rrpoi}). The main advantage of this deformation is an
alteration of the hamiltonian without an alteration of the
$R$-matrix algebra.

Let us introduce two matrices
 \ben
 F(u)&=&
 2u (l, g) g_3^{-1}
 \left(\begin{array}{cc}1&0\\0&0\end{array}\right)
 =2u J_3
 \left(\begin{array}{cc}D^{-1/2}&0\\0&0\end{array}\right)
 \,, \nn\\
 \ll{dop}\\
 T(u)&=&T_0(u)+F(u)\,. \nn
 \en
Matrix $T(u)$ is an additive deformation of the matrix $T_0(u)$
and they are coupled by certain deformation of the Sklyanin
brackets (\ref{rrpoi}). In the Sec.\ \ref{sec:l7} the similar dynamical
deformations will be considered for the Toda lattice associated
with the Lie algebra $\SG_2$.

 \begin{theorem}
For an arbitrary magnitude of the Casimir operator $J_3=(l, g)$
matrices $T_0(u)$ (\ref{vop}) and $T(u)$ (\ref{dop}) obey the
following relations
 \ben
 \left\{\,\on{T_0}{1}(u), \on{T_0}{2}(v)\,\right\}&=&
 \left[\,r(u-v), \,\on{T}{1}(u)\on{T}{2}(v)\,\right]=\ll{rspoi}\\
 &=&\left[\,r(u-v), \,\on{T_0}{1}(u)\on{T_0}{2}(v)\,\right]+W(u, v, l_j, g_j)
 \,,\nn\\
 \nn\\
 \left\{\,\on{T}{1}(u), \on{T}{2}(v)\,\right\}&
 =&\left[\,r(u-v), \,\on{T}{1}(u)
 \on{T}{2}(v)\,\right]
 +\ll{sspoi}\\
 &+&\on{T}{1}(u) s_2 \on{T}{1}(v)-\on{T}{1}(v) s_2 \on{T}{1}(u)
 +\on{T}{2}(u) s_1 \on{T}{2}(v)-\on{T}{2}(v) s_1 \on{T}{2}(u)\,.
 \nn
 \en
 The corresponding matrices have the form
 \ben
 W&=&\left[r(u-v), \on{F}{1}(u)\on{T}{2}(v)+\on{T}{1}(u)\on{F}{2}(v)\right]
 \,, \nn\\
 r&=&\dfrac{\eta P}{u-v}\,=
 \dfrac{2i}{u-v}\sum_{i=1}^3 \s_i\otimes \s_i\,,\qquad
 \eta=2i \ll{r}\\
 \nn\\
 s_1&=& \dfrac{-\eta u(l, g)}{4g_3^3\,(u-v)}\,(I+\s_3)
 \otimes(I-\s_3)=
 \dfrac{-\eta uJ_3}{g_3^3\,(u-v)}\,
 \left(\begin{array}{cccc}
 0&0&0&0\\0&1&0&0\\0&0&0&0\\0&0&0&0\end{array}\right)
 \,, \ll{s1}\\ \nn\\
 s_2(u,v)&=&P\,s_1(v,u)\,P\,,\ll{s2}
  \en
where $\s_i$ are Pauli matrices and $P$ is a permutation operator
of auxiliary spaces.
\end{theorem}
The proof is a direct but lengthy computation.

According by   \cc{ft87} the Lax representation is constructed
by using deformed $R$-matrix structure
 \bq
 \dot{T}_0(u)=\{H, T(u)\}=\left[M(u), T(u)\right]\,. \ll{laxneu}
 \eq
Matrix $M(u)$ is derived from the algebraic relations
(\ref{rspoi}) using the definition of the hamiltonian
\bq
 H=\Phi_v\,[\tr\,T_0(v)]=\Phi_v[t_0(v)]\,\qquad
 \hbox{ with }\left. \Phi_v[z(v)]\equiv z(v)\right|_{v=0}\,.
 \eq
 This matrix equal to
 \ben
 M(u)&=&-\Phi_v \tr_2\left[ I\otimes T(v)\cdot r(u-v)\,
 \right]\nn\\
 \nn\\
 &=&\left[\,\dfrac{-\eta}{u-v}\left(\begin{array}{cc}
 A(v)-D(v)&B(v)\\C(v)&D(v)-A(v)\end{array}\right)\,
 \right]_{v=0}\,. \ll{m1}
 \en
Here $A, B, C$ and $D$ are entries of the $T(u)$ (\ref{vop}) and $tr_2$
means trace in the second auxiliary space.
More precisely,
 \ben
 \dot{T}_0(u)&=&\left[M(u), T(u)\right]=\left[M(u), T_0(u)\right]+N\,,\nn\\
 \nn\\
 N&=&-4J_3
 \left(\begin{array}{cc}
 0&-l_+\\l_-&0\end{array}\right)\,.\nn
 \en
It is either a Lax pair at the level $J_3=0$ or a Lax triad for
an arbitrary magnitude of $J_3$.  This Lax representation is a
compatibility condition for the following linear problems
\ben
&&T(u)\,\varphi(u)-\l\varphi(u)=\psi(u)\,, \nn\\
\ll{abf}\\
&&\dfrac{d \varphi(u)}{dt}+M(u)\,\varphi(u)=0\,, \qquad
\dfrac{d \psi(u)}{dt}+M(u)\,\psi(u)=N(u)\,\varphi(u)\,, \nn
\en
where $\varphi$ is so-called Baker-Akhiezer function at the level
$J_3=(l. g)=0$.

Now the invariants are trace of $T_0(u)$,
which remains a generating function of the integrals of motion
(\ref{t0}), and determinant of $T(u)$:
 \ben
 t_0(u)=\hbox{tr}\,T_0(u)&=&u^2-2um-H\,, \nn\\
 \Delta(u)={\det}\,T(u)&=&u^2J_2\,. \ll{d}
 \en
The dual determinant of $T_0(u)$ and trace of $T(u)$ are the
dynamical variables:
 \ben
 \Delta_0(u)={\det}\,T_0(u)&=&\Delta(u)-2J_3\,g_3\,u\,, \ll{dd}\\
 t(u)=\hbox{tr}\,T(u)&=&t_0(u)+\dfrac{2J_3u}{g_3}\,. \nn
 \en

So, we obtained some starting point for a machinery of the
inverse scattering method. Below we introduce the quantum
counterpart of the presented deformation and consider a Lax
representation for the Goryachev-Chaplygin top. Then we will
try to apply the standard scheme related to the reflection
equations \cite{skl88}.


\section{Quantum axially symmetric Neumann's system}
\ll{sec:l4}
\setcounter{equation}{0}

Let variables $l_i, g_i, ~i=1, 2, 3$ be generators
 of the Lie algebra $e(3)$ with commutator relations:
 \ben
 &&\bigl[ l_i\,, l_j\,\bigr]= -i\eta\e_{ijk}\,l_k\,,
 \qquad \bigl[ l_i\,, g_j\,\bigr]= -i\eta\e_{ijk}\,g_k\,,\nn\\
 &&\bigl[ g_i\,, g_j\,\bigr]= 0\,, \qquad i, j=1, 2, 3.
 \nn
 \en
The quantum operator $T_0(u)$ related to a classical monodromy
matrix (\ref{vop}) has the form
 \bq
 T_0\,(u)=
 \left(\begin{array}{cc}
 u^2-2l_3\,u - l_1^2-l_2^2-\dfrac14~~~&
 i(g_+\,u-\dfrac12\{g_3\,, l_+\,\})\\
 i(g_-\,u-\dfrac12\{g_3\,, l_-\,\})&
 g_3^2\end{array}\right)\,,
 \ll{qvop}
 \eq
here braces $\{, \}$ mean an anticommutator.  Operator $T_0(u)$
(\ref{qvop}) at the level $J_3=(l, g)=0$ obeys the FCR
(\ref{fcr}) with the $R$-matrix of the $XXX$ type $R(u)=u+i\eta
P$ \cite{kuzts89}.

Introduce two additional matrices
\bq
F(u)=\dfrac{(u-i\eta) J_3}{g_3}
\left(\begin{array}{cc}1&0\\0&0\end{array}\right)\,, \qquad
T(u)=T_0(u)+F(u)\,. \ll{qdop}
\eq
Below the spectral parameter $u$ in the
$F(u)$ is always shifted by the constant $i\eta$.

\begin{theorem}
By the arbitrary values of the Casimir operator $J_3=(l, g)$
operator $T_0(u)$ (\ref{qvop}) obeys the following deformation of
the FCR (\ref{fcr})
\ben
 &&R(u-v)\,{\on{T_0}1}(u) \,{\on{T_0}2}(v) -
 {\on{T_0}2}(v) \,{\on{T_0}1}(u) R(u-v)=W(u, v, l_j, g_j)
 \ll{dfcr} \\
 &&W(u, v, l_j, g_j)=
 \left[\,{\on{F}1}(u)\,{\on{T_0}2}(v) +
 {\on{T_0}1}(u) \,{\on{F}2}(v)\,, \,R(u-v)\right]\,, \nn
 \en
 where $[, ]$ stands for a matrix commutator.
 \end{theorem}
The proof is a straightforward calculation.

The deformed FCR (\ref{dfcr}) assumes another forms:
 \ben
 &&R(u-v)\left({\on{T_0}1}\,{\on{T_0}2} +
 {\on{F}1}\,{\on{T_0}2} +{\on{T_0}1} \,{\on{F}2}\right)
 =
 \left( {\on{T_0}2} \,{\on{T_0}1} +
 {\on{F}1}\,{\on{T_0}2} +{\on{T_0}1} \,{\on{F}2}\right)
 R(u-v)\,, \nn\\
 &&\nn\\
 &&R(u-v)\,{\on{T}1} \,{\on{T}2} -
 {\on{T}2} \,{\on{T}1} R(u-v)=
 (\left[{\on{F}1}, {\on{T_0}2}\right]+
 \left[{\on{T_0}1} , {\on{F}2}\right])R(u-v)\,, \nn
 \en
(for the sake of brevity we have omitted the arguments $u,v$ in
the last two formulas).

Generating function of the quantum integrals of motions is the
trace of $T_0(u)$
\[t(u)=\tr\,T_0(u)\,, \qquad [t(u), t(v)]=0\,.\]
The quantum determinant $\qdet\,T(u)$ is the central element now
\bq
\Delta(u)\equiv \qdet \,T(u)=(u^2-1/4)J_ 2\,.
\eq

By using the deformation of the FCR (\ref{dfcr})
the quantum Kowalewski top will considered in the Sec.\ \ref{sec:l8}.


\section{The Goryachev-Chaplygin top}
\ll{sec:l5}
\setcounter{equation}{0}
The axially symmetric Neumann's system related to the
two-particle Toda lattice associated to the root system $\SA_2$
\cite{kuzts89}. It is well known that the Goryachev-Chaplygin
top (GCT) related to the three-particle Toda lattice associated
to the root system $\SA_3$ \cite{bvm87}.  Now we present
relations between the corresponding Lax representations.

The Goryachev-Chaplygin top (GCT)  represents a
 symmetric top in a constant homogeneous field with the principal
momenta of inertia satisfying $I_1\,:I_2\,:I_3=1:1:1/4$ and the
center of mass located in equatorial plane. hamiltonian of the
GCT is
\bq
J_1=H=\dfrac12\left(l_1^2+l_2^2+4l_3^2\right)-g_1\,.
\ll{hgc}
\eq
It is completely integrable in the one-parameter subset of
orbits ${\cal O}$ ($J_2=(g, g)=a^2$ and $J_3=(l, g)=0$) in
$e(3)^*$.

The GCT has been investigated in quantum inverse scattering
method by Sklyanin \cite{skl85} and generalized in   \cc{kk87a}.
It was a starting point for these investigations.

For construction of the monodromy matrix for the GCT we recall
basic results of   \cc{ts90}.
\begin{lemma}
Let the monodromy matrix $T(u)\in Y({\sf
g})\otimes\hbox{End}({\bC}^2)$ obeys the FCR with $R$-matrix of
the $XXX$ type. If exist such element $K\in Y({\sf g})$ that
\ben
&& [K(u), A(u)]=[K(u), D(u)]=0\,, \nn\\
&&[K(u), B(u)]=\eta B(u)\,, \quad [K(u), C(u)]=-\eta C(u)\,,\nn
\en
then the monodromy matrix
\bq
T_1(u)=
\left(\begin{array}{cc} u-p+K& \b e^{iq}\\
\g e^{-iq}&0\end{array}\right)
\left(\begin{array}{cc} A(u)& e^{iq} B(u)\\
e^{-iq}C(u)&D(u)\end{array}\right)\in
Y({\sf g}\oplus {\sf w})\otimes\hbox{End}({\bC}^2)\,,
\ll{nqum}
\eq
where generators of ${\sf w}$ are $[p, q]=-i\eta$
 and $\b, \g\in {\bC}$,
obeys the FCR as well.
\end{lemma}
The proof consists of in the fact that the two matrices
in the product (\ref{nqum}) obey the FCR (\ref{fcr}) with
one $R$-matrix and their entries mutually commute.

Definition of the monodromy matrix $T_1(u)$ (\ref{nqum}) can be
rewritten in other form
\bq
T_1(u)=\left(\begin{array}{cc} e^{-iq/2}& 0\\
0&e^{iq/2}\end{array}\right)
\left(\begin{array}{cc} u-p+K& \b\\
\g&0\end{array}\right)
\left(\begin{array}{cc} A(u)& B(u)\\
C(u)&D(u)\end{array}\right)
\left(\begin{array}{cc} e^{iq/2}& 0\\
0&e^{-iq/2}\end{array}\right)\,,
\nn
\eq
which we can consider as some kind of gauge transformation
of standard rule in quantum inverse scattering method \cite{ft87}.

\begin{lemma}
Let the matrix $T(u)\in Y({\sf g})\otimes\hbox{End}({\bC}^2)$
is a finite-dimensional irreducible representation of the
algebra of monodromy matrices (\ref{fcr}), which is polynomial
of spectral parameter $u$.
If the entries of the matrix $T(u)$ have the following
asymptotic behavior
\ben
&A\,(u)= u^N-a_1\,u^{N-1}+a_2\,u^{N-2}+\ldots\,, \qquad
&B\,(u)= b_1\,u^{N-1} +\ldots\,, \nn\\
\ll{asymp}\\
&C\,(u)= c_1\,u^{N-1} +\ldots\,, ~~~~~~~~~~~~~~~~\qquad
&D\,(u)= d_1\,u^{N-2} +\ldots\,, \nn
\en
then the element $K=a_1$ obeys the conditions of a Lemma 1.
Representations $T(u)$ and $T_1(u)$ (\ref{nqum}) are related to
the integrable systems with the following integrals of motion
\ben
&t(u)=\hbox{tr}\,T(u)=u^N-a_1\,u^{N-1}
+&(a_2+d_1)\,u^{N-2}+\ldots\,, \nn\\
&t_1(u)=\hbox{tr}\,T_1(u)=
u^{N+1}-pu^{N}+&(a_2-a_1^2-pa_1+\b c_1
+\g b_1)u^{N-1}+\ldots\,, \nn
\en
where operator $p$ has a continuous spectra.
\end{lemma}
For the proof we have to substitute the asymptotes (\ref{asymp})
to FCR (\ref{fcr}).

These Lemma's have been introduced to $R$-matrices
of $XXX$ and $XXZ$ types in   \cc{ts90}
by considering the classical and relativistic
Toda lattices in the Jacoby systems of coordinates.
By using these Lemma's  and the  matrix $T_0(u)$
(\ref{qvop}) we obtain the new monodromy matrix $T_1(u)$
with entries
\ben
A(u)=&&(u-p+2l_3)\left
(u^2-2l_3u-l_1^2-l_2^2-1/4-\dfrac{\mu^2-1/4}{g_3^2}\right)
\nn\\
&&+i\b[\,ug_- -\{g_3, l_-\}/2\,]
\,, \nn\\ \nn\\
B(u)=&&e^{iq}[i(u-p+2l_3)(\,ug_+-\{g_3, l_+\}/2\,)
+\b g_3^2]
\,,
\ll{egct}\\
\nn\\
C(u)
=&&e^{-iq}[\,u^2-2ul_3-l_1^2-l_2^2-1/4-\dfrac{\mu^2-1/4}{g_3^2}\,]\,,
\nn\\ \nn\\
D(u)=
&&i\g(\,ug_+-\{g_3, l_+\}/2\,)\,. \nn
\en
Matrix $T_1(u)$ at the level $\mu=1/2$
has been introduced in   \cc{skl85} and generalized for an
arbitrary magnitude of $\mu$ in   \cc{kk87a}.


\section{ The Kowalewski-Chaplygin-Goryachev top 
and reflection equation}
\ll{sec:l6}
\setcounter{equation}{0}

Consider representations $U(u)$ of twisted yangians related to the
reflection equation (\ref{re}--\ref{repoi}).  The algebra of
monodromy matrices (\ref{fcr}) and its classical counterpart
(\ref{repoi}) have two important automorphisms $T(u)\to
T^{\,\s}(u)$:
\bq
T^{\,a}(u)=\s_2 T(u)\s_2\approx (T^{-1})^{\,t}(u)\qquad
T^{\,i}(u)=\s_2 T^{\,t}(-u)\s_2\approx T^{-1}(-u)\,.
\ll{iso}
\eq
The symbol $T^{\,t}$ means a transposition matrix and symbols
$T^{\,a}$ and $T^{\,i}$ related to antipod map and to involution map,
respectively, in theory of quantum groups \cite{skl88}.

In the classical mechanics monodromy matrix $U(u)$ related to
reflection equation can be constructed as
\bq
U(u)=T_+(u)T_-(u)\,,
\ll{ucomp}\eq
with the matrices $T_\pm$ defined by
\ben
T_-(u)&=&T_1(u)K_-(u)T_1^{\,i}(u)\,, \nn\\
T^{\,t}_+(u)&=&T_2^{\,t}(u)K^{\,t}_+(u)T_2^{\,a}(-u)\,. \nn
\en
Here matrices $T_j(u), ~j=1, 2$ obey the Sklyanin brackets
(\ref{rrpoi}) with some matrix $r(u)$, $K_\pm(u)$ are known
solutions to reflection equation (\ref{repoi}) with same matrix
$r(u)$ and $s(u+v)=r(u+v)$. The matrices
$T_\pm(u)$ (\ref{ucomp}) obey the classical reflection equation
as well \cite{skl88}.

Consider a simple ${\bC}-$number solutions of RE
(\ref{re}) $K_\pm(u)$, which correspond to various boundary
conditions for integrable systems \cite{skl88}.  For the rational
$R$-matrix of $XXX$ type matrices $K_\pm(u)$ are given by
\ben
K_-(u, \a_1, \b_1, \g_1)&=&
\left(\begin{array}{cc} \a_1+{u}\b_1& u\\
u\g_1&\a_1-{u}\b_1\end{array}\right)\,, \nn\\
\ll{kpm}\\
K_+(u, \a_2, \b_2, \g_2)&=&
\left(\begin{array}{cc} -\a_2+{u}\b_2& -u\g_2\\
-u&-\a_2-{u}\b_2\end{array}\right)\,. \nn
\en
For instance, the boundary matrices $K_\pm$ (\ref{kpm}) are
applied to construct the monodromy matrices for the Toda lattices
associated to the $\SB_n$ and $\SC_n$ root systems \cite{skl88}.

By taking matrix $T_1(u-\g)$ (\ref{fneu}) with the gyrostat
parameter $\g$ and matrices $K_\pm(u)$ (\ref{kpm}) we construct
the new monodromy matrix $U(u)$ as (\ref{ucomp}) ($T_2(u)=I$).
At the level $J_3=0$  we get the completely integrable system
with integrals of motion defined by the trace of $U(u)$.  The
hamiltonian of this system is equal to
\ben
H&=& l_+ l_-+2 l_3^2-i(\a_1g_+- \a_2 g_-)+
\dfrac12(\g_1 g_+^2 -\g_2g_-^2)
+\dfrac{\mu}{g_3^2}\nn\\
\ll{hamg}\\
&-&i g_3 (\b_1 l_+-\b_2 l_-)
-2 i l_3(\b_2 g_-- \b_1 g_+)
+\g(2l_3+i\b_1 g_+-i\b_2 g_-)
\,, \nn
\en
If $\g=\b_1=\b_2=0$ integrable system with the
hamiltonian (\ref{hamg}) can be identified with the
Kowalewski-Chaplygin-Goryachev top \cite{kuzts89}.

In quantum mechanics operators $T_\pm(u)$ are
\ben
T_-(u)&=&T(u-\g)K_-(u-\dfrac{\eta}2)\s_2
T^{\,t}(-u-\g)\s_2\,, \ll{umq}\\
T_+(u)&=&K_+(u+\dfrac{\eta}2)\,. \nn
\en
Here operators $T(u)$ and $K_\pm(u)$ are representations
of monodromy matrix algebras related to FCR (\ref{fcr}) and to RE
(\ref{re}) with quantum matrices $R(u)$ and $S(u+v)=R(u+v-\eta)$,
respectively.

Thus, we describe the eight parameters family ($\a_j\,, \b_j\,,
\g_j$ with $j=1, 2$; $\mu\,, \g$) of completely integrable at the
level $J_3=(l, g)=0$ systems on the Lie algebra $e(3)$ in quantum
inverse scattering method (for comparison see   \cc{bog84}).


\section{Toda lattice associated to the Lie algebra
$\SG_2$}
\ll{sec:l7}
\setcounter{equation}{0}

The family of tops introduced in the previous section can be
associated with the two-particle Toda lattices related to the
root system $\SB\SC_2$ \cite{kuzts89}.  The remaining non-trivial
two-root system is $\SG_2$.  We hope that consideration of the
monodromy matrix in the two dimensional auxiliary space for the
corresponding Toda lattice gives us some background for a search
of the non-standard Lax representation for the KT.

The group $\SG_{2}$ is of rank two and dimension 14 and it has
two simple roots $\a_1$ and $\a_2$.
The Weyl group of $\SG_{2}$ is the permutation group of order 3
with inversion, generated by $\tau_{1}$ and $\tau_{2}$,
\bq
\tau_{1} :\, (a_{1},a_{2})\to (a_{2}-a_{1},a_{2})\,, \qquad
\tau_{2} :\, (a_{1},a_{2})\to (-a_{1},a_{2}-3a_{1})\,.
\ll{we}
\eq
The root system is easier to describe in the standard basis in
a large space $\bR^3$. We will use tree pairs of canonically
conjugate variables $(q_j,p_j)$ with a linear constraint $\sum
q_j=\sum p_j=0$. The non-constrainted system can be obtained by
using the following canonical transformation
\ben
&&{q}_1\to \sqrt3 q_1+q_2+\dfrac{q_3}3\,,\qquad
{q}_2\to -2 q_2+\dfrac{q_3}3\,,\qquad
{q}_3\to - \sqrt3 q_1 + q_2+\dfrac{q_3}3\,,\nn\\ \nn\\
&&{p}_1\to \dfrac{p_1}{2\sqrt3}+\dfrac{p_2}6+p_3\,,\qquad
{p}_2\to-\dfrac{p_2}3+p_3\,,\qquad
{p}_3\to-\dfrac{p_1}{2\sqrt3}+\dfrac{p_2}6+p_3\,,\nn\\ \nn
\en
which transforms the corresponding hamiltonian to the natural form.

According by   \cc{avm80} the Lax representation is equal to
\ben
\SL&=& \left(\begin{array}{ccccccc}
-\b_3&-\a_1& 0&0&0&0&0\\
-1&-\b_2 &-\a_2&0&0&0&0\\
0&-1&-\b_1&-2 \a_1&0&0&0\\
0&0&-1&0&2 \a_1&0&0\\
0&0&0&1&\b_1&\a_2&0\\
0&0&0&0&1&\b_2 &\a_1\\
0&0&0&0&0&1&\b_3\end{array}\right)\ll{gl}\\
\nn\\
{\rm where}\qquad&&\b_1={p}_3-{p}_1\,,\qquad
\b_2={p}_1-{p}_2\,,\qquad
\b_3=p_3-p_2\,,\nn\\ \nn\\
&&\a_1=e^{a_1}=e^{{q}_1-{q}_2}\,,\qquad
\a_2=3e^{a_2}=3e^{-2{q}_1+{q}_2+{q}_3}\,,\nn
\en
It is a three diagonal matrix and, therefore, we can easy
obtain the corresponding monodromy matrix in the two
dimensional auxiliary space \cite{ft87}.

Let us define three matrices $L_j(u)$
\ben
L_1&=& \left(\begin{array}{cc}
u-{p}_1-{p}_2&- e^{{q}_1 -{q}_2}\\
 e^{-{q}_1 +{q}_2}&0\end{array}\right)\,,\qquad
\L_2= \left(\begin{array}{cc}
u-{p}_3-{p}_1&-3e^{{q}_3 -{q}_1}\\
 3e^{-{q}_3 +{q}_1}&0\end{array}\right)\,,\nn\\
 \nn\\
L_3&=& \left(\begin{array}{cc}
u-{p}_3-{p}_2&-\dfrac13 e^{{q}_3 -{q}_2}\\
 \dfrac13e^{-{q}_3 +{q}_2}&0\end{array}\right)\,, \ll{mg2}
\en
and two boundary matrices $T_\pm(u)$
\ben
K_-&=&\left(\begin{array}{cc}
  2&u\\0&2\end{array}\right)\,,\qquad
K_+=\left(\begin{array}{cc}
  0&0\\-u&0\end{array}\right)\,,\nn\\
\nn\\
T_-&=&L_1K_-L_1^\s\,,\qquad
T_+=L_3^\s K_+L_3\,.\ll{mgpm}
\en
The monodromy matrix for the open Toda lattice associated
to the algebra $\SG_2$ is equal to
\ben
U(u)&=&T_+(u)L_2(u)T_-(u)L_2^\s(u)\,,\nn\\
\ll{g22}\\
\tr U(u)&=&-u \det\,(\SL-uI)=
-u^8+h_1u^6+h_2u^4+h_3u^2\,,\nn
\en
where $h_j$ are integrals of motion.
This form for matrix $U(u)$ is a more symmetrical form for the
Weyl group (\ref{we}).  For the affine algebra $\SG_2^{(1)}$ we
have to substitute new matrix $T_+(u)$ into (\ref{g22})
\[
\te{T}_+=
\left(\begin{array}{cc}
 \dfrac{(u-p) e^q+(u+p)e^{-q}}3 & -\dfrac19( e^q-e^{-q} )^2\\
\\-(u^2-p^2)& \dfrac{(u+p)e^q+(u-p)e^{-q}}3\end{array}\right)\,,\]
where $p=p_3-p_2$ and $q=q_3-q_2$

The Poisson brackets relations for the matrices $L_k(u)$ (\ref{mg2})
have the following polylinear form
\ben
\{{\on{L}{1}}_j(u), {\on{L}{2}}_k(v)\}&=&
\d_{jk}[r(u-v), {\on{L}{1}}_j(u)\,{\on{L}{2}}_k(v)\,]+\nn\\
\ll{plr}\\
&+&(1-\d_{jk})(-1)^{2j-k}\,
\left([r_1,{\on{L}{1}}_j(u)]+[r_2,{\on{L}{2}}_k(v)]\right)\,,\nn\\
\nn\\
&&\qquad j\ge k\,,\quad j,k=1,2,3.\nn
\en
here $r(u-v)$ is a standard $R$-matrix of the $XXX$
type (\ref{r}) and the
independent from spectral parameter matrices $r_{1,2}$ are given by
\bq
r_1=-\dfrac14(I-\s_3)\otimes(I+\s_3)\,,
\qquad
r_2=Pr_1P\,,\ll{rmm}
\eq
(in comparison with (\ref{sspoi})). Locally ($j=k$) these
Poisson brackets relations are the standard Sklyanin brackets
and, therefore, matrices $T_\pm(u)$ (\ref{mgpm}) obey the
standard non-dynamical RE (\ref{re}).

By using this polylinear algebra and factorization (\ref{g22}) the
basic property of trace of a monodromy matrix $U(u)$
\[\{t(u),t(v)\}=0\,,\qquad t(u)=\tr U(u)\,,\]
can be easy proved.

If we assume that some representation $U(u)$ associated to the
root system $\SG_2$ can not be expanded on the simplest factors
as above, then we must introduce a more complicated dynamical
$R$-matrix structure.
Consider two matrices $T^{(1,2)}(u)$ defined by
\bq
 T^{(1)}(u)=L_2(u)L_1(u)\,,\qquad T^{(2)}(u)=L_3(u)L_2(u)L_1(u)\,.
\ll{mt2}
\eq
Their Poisson brackets relations are calculated from the
polylinear algebra (\ref{plr}) and we can prove that we can not
close these relations at the quadratic $R$-matrix algebra by
using only the $\bC$-number $R$-matrices.  These Poisson brackets
relations have the form of the deformed Sklyanin brackets
(\ref{sspoi}) with the dynamical matrices $s_{k}=\a_k(p,q)
r_k\,,\quad k=1,2$ (\ref{rmm}).  Here expressions for the
dynamical coefficients $\a_k(p,q)$ are simply recovered from
(\ref{plr}).

Matrices $T^{(j)}$ have one common property for their traces
\ben
&&t^{(j)}(u)=\tr T^{(j)}(u)=u^n+h_1u^{n-1}+
h_2u^{n-2}+h_3\,,\quad j=1,2\quad n=j+1\,,\nn\\
&&\{t^{(1)}(u),t^{(1)}(v)\}=u-v\,,\qquad
\{t^{(2)}(u),t^{(2)}(v)\}=(u-v)(uv+u+v+1)\,,\nn\\
&&\{h_i,h_k\}=1\,,\quad i> k\,,\quad i,k=1,2,3,\nn
\en
and, of course, these traces do not the generating functions of
integrals of motion.

The corresponding matrices $T_-^{(j)}(u)=T^{(j)}K_-T^{(j)\s}$
obey the dynamical deformations of the classical reflection
equation (\ref{rrpoi})
\bq
\left\{\,{\on{T}{1}}_-, {\on{T}{2}}_-\,\right\}=
\left[\,r(u-v), {\on{T}{1}}_-{\on{T}{2}}_-\,\right]+
{\on{T}{1}}_-\,r(u+v)\,{\on{T}{2}}_--
{\on{T}{2}}_-\,r(u+v)\,{\on{T}{1}}_-+W(u,v,p_j,q_j)\,,\ll{dre}
\eq
where $W(u,v,p_j,q_j)$ is matrix-function of spectral
parameters and of dynamical variables.
We choose the simplest form for the dynamical deformations of
the RE (\ref{dre}). Of course, matrix $W$ can be presented as
the various combinations of $T_-^{(j)}$ and proper dynamical
$R$-matrices. May be, these combinations wiil more deeply
reflect a structure of the Weyl group of $\SG_2$ (\ref{we}).

We have to emphasize here that matrices $T_-^{(j)}$
relate to at most then another factorization of the monodromy
matrix $U(u)$ (\ref{g22})
\[U(u)=T_+(u)T_-^{(1)}(u)=K_+(u)T_-^{(2)}(u)\,.\]
Recall that the monodromy matrix $T_0(u)$ (\ref{vop}) for the
Neumann's system is closely connected to the matrix $T(u)$
(\ref{todafact}) for the Toda lattice. However, the structure
of the phase spaces are different \cite{bvm87}. Therefore, for
the Toda lattice $2\times 2$ matrix $T(u)$ are factorized on
the one-particles matrices (\ref{todafact}). The corresponding
matrix $T_0(u)$ in $e(3)^*$ can not be expanded on the simplest
factors. Then for the some integrable system associated with
the exceptional algebra $\SG_2$ we could get the monodromy matrix
$U(u)$ without the simplest expansion as (\ref{g22}).


\section{The Lax triad for the Kowalewski top}
\ll{sec:l8}
\setcounter{equation}{0}
Motivated by the previous example we will look the such
non-factorable monodromy matrix for the KT. Let boundary
matrices $K_\pm$ be
\bq
 K_+(u)=
 \left(\begin{array}{cc} -\a_2& 0\\
 -u&-\a_2\end{array}\right)\,, \quad\hbox{and}\quad
 K_-(u)=
 \left(\begin{array}{cc} \a_1& u\\
 0&\a_1\end{array}\right)\,, \ll{kkpm}
 \eq
(in comparison with (\ref{mgpm})).

The monodromy matrix $U(u)=K_+T_0(u)K_-T^{\,i}_0(u)=K_+T_-$, with
$T_0(u)$ given by (\ref{vop}), corresponds to the KT in the
one-parameter subset of orbits $\cal O$ ($J_2=(g,g)=a^2$ and
$J_3=(l, g)=0$).  Now we try to take up such additive deformation
that the deformed monodromy matrix is described the KT on whole
phase space $J_3\neq 0$.  Notice, that for the deformed Sklyanin
brackets (\ref{rspoi}) the maps $T(u)\to T^\s(u)$ (\ref{iso}) are
the automorphisms as well.

 \begin{theorem}
For the Kowalewski top on the Lie algebra $e(3)$ with
integrals of motion
\ben
J_1&=&H=(l_+ l_-+2 l_3^2)
- i(\a_1 g_++\a_2 g_-)
\,, \nn\\
J_2&=&(g, g)=a^2\,, \nn\\
J_3&=&(l, g)\,, \ll{kowint}\\
J_4&=&k_+k_-=(l_+^2+2i\a_2 g_+)(l_-^2 + 2i \a_1 g_-)\,, \nn
\en
the monodromy matrix $U(u)$ is given by
\bq
U(u)=K_+(u)T_-(u)\,. \ll{ukow}
\eq Here
\ben
T_-(u)&=&
\left(\begin{array}{cc}A(u)&B(u)\\C(u)&A(-u)
\end{array}\right)=
T_0(u)K_-(u)T^{\,i}_0(u)+G_-=
\nn\\
\ll{tm}\\
&=&T_0(u)K_-(u)T^{\,i}_0(u)+uJ_3
\left(\begin{array}{cc}il_-u+\a_1g_3&2i\a_1l_+\\
0&il_-u-\a_1g_3\end{array}\right)
\,. \nn
\en
More precisely,
\ben
A(u)&=&-i u (u^3 g_-+(g_3 l_--2 l_3 g_-) u^2
+(-i a^2 \a_1 +g_+ k_-) u
-g_3 l_+ k_-)\,, \nn\\
\nn\\
B(u)&=&u^5-2 (l_+l_-+2l_3^2-i\a_1 g_+) u^3+l_+^2 k_-u\,, \nn\\
\nn\\
C(u)&=&-u^3 g_-^2+g_3^2 k_-u\nn
\en

Matrix $U(u)$ (\ref{ukow}) obeys the Lax
representation in the form
\ben
\dfrac{dU(u)}{dt}&=&
\left[\,K_+M_-\,, U(u)\,\right]+K_+N_-\,, \ll{triadkow}\\
 \nn\\
M_-&=&2i
\left(\begin{array}{cc} -ig_-&-1\\0&ig_-\end{array}\right)\,,
\ll{mkow}\\ \nn\\
N_-&=&uJ_3
\left(\begin{array}{cc}
3l_-u^2 -l_+k_-& 0\\
2ig_3 k_-&-3l_- u^2 +l_+k_-)
\end{array}\right)\,. \ll{gkow}
\en

Trace of monodromy matrix $U(u)$ is a generating functions of
integrals of motion
\bq
t(u)=-u^6+2u^4 J_1-u^2\left(J_4+2\a_1\a_2 J_2\right)\,,
\eq
\end{theorem}

Matrix $T_-(u)$ has a typical to RE (\ref{repoi}) symmetry
property $T_-^{\,i}(u)=T_-(u)\,,$ which relates to the
involutions on the phase space.  It reflects the corresponding
symmetry of the Jacobian of spectral curve defined by the
$4\times 4 $ matrix $L(\l)$ (\ref{4lax}).

For calculation of the corresponding Poisson structure
we express $T_-(u)$ through matrices $T_0(u)$
and $F(u)$ (\ref{vop}--\ref{dop}), which have the known Poisson
structure (\ref{rspoi}).  Assume
\bq
s_a=
\left(\begin{array}{cc} 1& 0\\
0&0\end{array}\right)=
\dfrac{I+\s_3}2\,, \quad
s_b=
\left(\begin{array}{cc} 0& 0\\
0&1\end{array}\right)
=\dfrac{I-\s_3}2\,,
\eq
then the monodromy matrix $T_-(u)$ (\ref{tm}) is equal to
\ben
T_-(u)&=&T_0(u-\g) K_-(u)T^{\,i}_0(u-\g)+\nn\\
\ll{ttm} \\
&+&
\left[T_0(u-\g) s_b+\dfrac12
s_b(T_0(u-\g)-T^{\,i}_0(u+\g)\s_2)s_a\,\right]K_-(u)
F^{\,i}(u-\g)\nn\\
\nn\\
&+&F(u-\g)K_-(u)\left[\,s_aT^{\,i}_0(u-\g) +\dfrac12
s_b(T^{\,i}_0(u-\g)-T_0(u+\g)\,)s_a\,\right]\,.
\nn
\en
Here
\[T_0^{\,i}(u-\g)=\s_2T^{\,t}_0(-u-\g)\s_2=
\s_2T^{\,t}_0\left(-(u+\g)\right)\s_2\]
and we introduce a shift of the spectral parameter $u\to u-\g$ for
description of the Kowalewski gyrostat.
Then the trace of $U(u)$ is a generating function of
integrals of motion for the Kowalewski gyrostat
\ben
t(u)&=&-u^6+2u^4(J_1+\g^2)-u^2\left(J_4
+2\a_1\a_2 J_2-
\g^2(2 J_1+\g^2)\right)-2\g\a_1\a_2J_2\,, \nn\\
\nn\\
J_1&=&H= (l_+ l_-+2 l_3^2+2\g l_3)
- i(\a_1 g_++\a_2 g_-)
\,, \nn\\ \nn\\
J_4&=&k_+k_-
-4\g(l_3+\g)l_+l_-
-4i\g\,g_3(\a_1 l_++\a_2 l_-)
\,. \nn
\en

Next by using factorization (\ref{ttm}) and the deformed Sklyanin
brackets (\ref{rspoi}), we can prove that matrix $T_-(u)$ obeys
the following deformations of the classical reflection equation
\bq
\left\{\,{\on{T}{1}}_-, {\on{T}{2}}_-\,\right\}=
\left[\,r(u-v), {\on{T}{1}}_-{\on{T}{2}}_-\,\right]+
{\on{T}{1}}_-\,r(u+v)\,{\on{T}{2}}_--
{\on{T}{2}}_-\,r(u+v)\,{\on{T}{1}}_-+W(u, v, l_j, g_j)\,.
\ll{respoi}
\eq
For the KT ($\g=0)$ the dynamical matrix $W(u, u, l_j, g_j)$
is given by
\bq
W(u, v, l_j, g_j)=uvJ_3
\left(\begin{array}{cccc}
a(u, v)&b(u, v)&-b(v, u)&0\\
c(u, v)&a(u, -v)&d(u, v)&-b(-v, u)\\
-c(v, u)&-d(v, u)&a(-u, v)&b(-u, v)\\
0&-c(-v, u)&c(-u, v)&a(-u, -v)
\end{array}\right)\,,
\ll{wcl}
\eq
 where
\ben
a(u, v)&=&ik_-\left(g_3(u^2-v^2)-2g_+l_-(u-v)\right)\,, \nn\\
b(u, v)&=&6l_-u^2v^2-2l_+k_-(u^2-4l_3u+l_+l_-+v^2)\,, \nn\\
c(u, v)&=&2g_3k_-(g_3l_--2ug_-)\,, \ll{entrw}\\
d(u, v)&=&4ig_3k_-(u^2+l_+l_-)\,. \nn
\en
Dynamical matrix $W(u, v, l_j, g_j)$ can be expressed in the terms
of the matrices $T_0$, $F$ and $R$ by using the representation
(\ref{ttm}) and the deformed Sklyanin brackets (\ref{rspoi}) as well.

The Lax representation (\ref{triadkow}) are constructed from this
deformed $R$-matrix brackets (\ref{respoi}) according to
  \cc{ft87}, as for the Neumann's system (\ref{laxneu}).
The Lax triad has an additive freedom for the matrix $M$, we
always can pass from the Lax triad $(U, M, N)$ to triad $(U, M+M_1,
N+[L, M_1])$.  Our choice (\ref{mkow}) is fixed by the Lax pair at
the level $J_3=(l. g)=0$, which relates to a pure reflection
equations (\ref{repoi}).

We understand, that we present a few artificial construction in
comparison with the construction of the Lax representation on
loop algebras by Reyman and Semenov-Tian-Shansky \cite{rs87}.
However, we have some new positive properties of the proposed Lax
representation.

Under the following transformation of universal enveloping algebras and
of definition of matrix $T_-(u)$
\bq
k_-\to k_-\pm 1/4 \quad\hbox{and}\quad
T_-(u)\to T_-(u)\mp ik_-u^2\sigma_3/4\,,
\label{so4}
\eq
the new monodromy matrix $U(u)=K_+(u)T_-(u)$ describes the Kowalewski top
on the Lie algebras $so(4)$ and $so(3, 1)$ with integrals of motion
introduced in   \cc{kom87}.

Monodromy matrix for the quantum Kowalewski top can be obtained
according to a general scheme \cite{skl88}.  Substituting into
definition of monodromy matrix (\ref{ttm}) quantum operators
$T_0(u)$ and $F(u)$ (\ref{qvop}--\ref{qdop}) and boundary
matrices $K_\pm(u\pm \eta/2)$, we obtain the quantum
$U$-operator for the Kowalewski top.  It means that trace of
this $U$-operator is a generating function of true integrals of
motion in the quantum mechanics
\ben
t(u)&\equiv&\tr\,U(u)\,, \qquad [t(u), t(v)]=0\,, \nn\\
J_1&=& H=l_1^2+ l_2^2+2 l_3^2
- i\a_1 g_+-i\a_2 g_-
\,, \ll{kowhq}\\
J_4&=&\dfrac12\{k_+, k_-\}+2\eta^2\{l_+, l_-\}\,. \nn
\en
Here operators $k_\pm$ have been defined in (\ref{kowint}) and
braces $\{, \}$ mean anticommutator of quantum operators.
Operator $T_-(u)$ obeys the deformation of quantum reflection
equation
\[
R(u-v)\,{\on{T_-}1}(u)\,R(u+v-\eta)\,{\on{T_-}2}(v)-
{\on{T_-}2}(v)\,R(u+v-\eta)\, {\on{T_-}1}(u)\,R(u-v)=W\,.  \]
Here quantum matrix $W(u, v, l_j, g_j)$ is obtained by
application of the deformed FCR (\ref{dfcr}) to the operator
$T_-(u)$ defined by (\ref{ttm}).

\section{Conclusions}
We present a monodromy matrix on $2\times 2$ auxiliary space for
classical and quantum Kowalewski top. This matrix relates with
the additively deformed reflection equation (\ref{respoi}).
Deformations $W(u, v, l_j, g_j)$ (\ref{wcl}) depends on the
spectral parameters and the dynamical variables. The dynamical
deformation of quadratic $R$-matrix algebras can be considered as
analog of dynamical $r$-matrices for linear $R$-matrix algebras
\cite{ts94,krw95}.

However this complication of inverse scattering method will be
justified if it allows to describe a quite wide set of integrable
systems, as for linear dynamical $R$-matrix, or to solve
completely such nontrivial system as the Kowalewski top. We hope
to obtain such confirmations in the forthcoming publications.


\section{Acknowledgements}
The author would like to express his gratitude to S. Rauch-Wojciechowski
for valuable discussions and to Department of Mathematics of
Linkoping University for warm hospitality.
This work was supported by Swedish Royal Academy of Science grant,
project N1314 for collaboration with Russian Academy of Science,
and RFBR grant.

\end{document}